\documentclass[amssymb,aps,showpacs,showkeys, twocolumn,prl]{revtex4-1}

\usepackage{latexsym}
\usepackage{graphicx,color}
\usepackage{graphicx}
\usepackage{amsmath,amssymb}
\usepackage{amsmath}
\usepackage[T1]{fontenc}
\usepackage{float} 
\usepackage[hang,small,up]{caption}
\usepackage{subfigure}  
\usepackage{lineno}

\begin{document}

\title{The fractal geometry of growth: fluctuation-dissipation theorem  and hidden symmetry}

\author{Petrus H. R. dos Anjos$^{1}$}

\author{M\'{a}rcio S. Gomes-Filho$^{2}$} 

\author{Washington S. Alves$^{2}$}

\author{David L. Azevedo$^{2}$}

\author{Fernando A. Oliveira$^{2,3}$}
\email{faooliveira@gmail.com}

\affiliation{Instituto de F\'{i}sica, Universidade Federal de Catal\~{a}o, CEP 75704-020 - Catal\~{a}o, GO - Brasil$^{1}$}

\affiliation{Instituto de F\'{i}sica, Universidade de Bras\'{i}lia, Bras\'{i}lia-DF, Brazil$^{2}$}
\affiliation{Instituto de F\'{i}sica, Universidade Federal da Bahia, Campus Universit\'{a}rio da Federa\c{c}\~{a}o,
 Rua Bar\~{a}o de Jeremoabo s/n, 40170-115, Salvador-BA, Brazil $^{3}$ }

\begin{abstract}
Growth in crystals can  be  { usually } described by  field equations such as the Kardar-Parisi-Zhang (KPZ) equation. While the crystalline structure can be characterized by Euclidean geometry with its peculiar symmetries, the growth dynamics creates a fractal structure at the interface of a crystal and its growth medium, which in turn determines the growth.  Recent   work (The KPZ exponents for the 2+ 1 dimensions, MS Gomes-Filho, ALA Penna, FA Oliveira;
\textit{Results in Physics}, 104435 (2021)) associated the fractal dimension of the interface with the growth exponents for KPZ, and provides explicit values for them. In this work we discuss how the fluctuations  and the responses to it are associated with this fractal geometry and the new hidden symmetry associated with the universality of the exponents.
\end{abstract}

\keywords{Non-equilibrium growth; Scaling laws; Stochastic analysis; Brownian motion; Noise; Symmetries; Fractals}

\maketitle

\section{ Introduction}

Symmetry and spontaneous symmetry breaking  are fundamental concepts to understanding nature, in particular to investigate the phases of matter. Since growth is one of the most ubiquitous phenomenon in nature, a question appears immediately ``What are the symmetries associated with growth?''  In order to  address  this question we shall look to the major growth  field equations. Two basic equations are
the Edwards Wilkinson (EW)~\cite{Edwards82}:
\begin{equation}
\label{EW}
\dfrac{\partial h(\vec{x},t)}{\partial t}=\nu \nabla^2 h(\vec{x},t)  + \xi(\vec{x},t),
\end{equation}
and the  Kardar-Parisi-Zhang  equation~\cite{Kardar86}:
\begin{equation}
\label{KPZ}
\dfrac{\partial h(\vec{x},t)}{\partial t}=\nu \nabla^2 h(\vec{x},t) +\dfrac{\lambda}{2}[\vec{\nabla}h(\vec{x},t)]^2 + \xi(\vec{x},t),
\end{equation}
where $h: \mathbb{R}^d \times \mathbb{R}^+ \rightarrow \mathbb{R}$ {denotes} the interface heights at point $\vec{x} \in \mathbb{R}^d$ at the time $t\geqslant0$. Since $h(\vec{x},t)$ displays different scale properties than those displayed by $\vec{x}$, we say that we have a $d+1$ dimensional space, in such way that the growth of a film will be in  $2+1$ space dimensions.
The parameters $\nu$ (surface tension) and $\lambda$ are related to the Laplacian smoothing and the tilt mechanism, respectively. The stochastic process is characterized by the zero mean white noise, $\xi(\vec{x},t)$,  with:
\begin{equation}
\label{xi}
\left\langle \xi(\vec{x},t) \xi(\vec{x}',t')\right\rangle = 2D\delta^{(d)}(\vec{x}-\vec{x}')\delta(t-t'),
\end{equation}
where $D$ is here the noise intensity. The above equation is sometimes {called} the Fluctuation-dissipation theorem (FDT). 
The EW equation were obtained~\cite{Edwards82,Barabasi95} considering the basic symmetries $\vec{x} \rightarrow \vec{x}+\vec{x}_0$,  $\vec{x} \rightarrow -\vec{x}$,   $t \rightarrow t+t_0$,  $h \rightarrow h+h_0$ and $h \rightarrow -h$, \textit{i.e.}  independence of the frame of reference. Note that the  symmetry $h \rightarrow -h$ is violated for KPZ due to the presence of the non-linear dependence on the
local slope $[\vec \nabla h]^2$. 
 The KPZ equation describes very well the dynamics of    some atomistic models such as the etching model~\cite{Mello01,Reis05,Almeida14,Rodrigues15,Alves16,Carrasco18}, and the Single-Step (SS) model~\cite{Krug92,Krug97,Derrida98,Meakin86,Daryaei20} {in the long wavelength limit}.  {For atomistic models we define our Euclidean space as a $d$-dimensional  hypercubic lattice within the region $\Omega \subset \mathbb{R}^d$, with volume} $V=L^d$, where $L$ is the lateral side.

Two quantities play an important role in growth, the average height, 
 $ \langle h(t) \rangle $, and the standard deviation
\begin{equation}
w(L,t)= \left[ \langle h^2(t) \rangle - \langle h(t) \rangle^2\right]^{1/2},
\label{wt}
\end{equation}
which is named as roughness or the surface width. {Here the average is taken over the space}.
The roughness is a very important physical
quantity, since many important phenomena  have been associated with it~\cite{Edwards82,Kardar86,Barabasi95,Mello01,Reis05,Almeida14,Rodrigues15,Alves16,Carrasco18,Krug92,Krug97,Derrida98,Meakin86,Daryaei20,Hansen00}.
For many growth processes, the roughness, $w(L,t)$, increases with time until reaches a saturated roughness $w_s$, i.e., $w(t \rightarrow \infty)=w_s$. We can summarize the time evolution of all regions as following~\cite{Barabasi95}: 
 \begin{equation}
\label{Sc1}
w(L,t)=
\begin{cases}
 ct^\beta , &\text{ if~~ } t <<t_\times\\
 w_s \propto L^\alpha, &\text{ if~~ } t >> t_\times,\\
\end{cases}
\end{equation}
with $t_{\times} \propto L^z$. The dynamical exponents satisfy the general scaling relation:
\begin{equation}
\label{z}
z=\frac{\alpha}{\beta}.
\end{equation}
The set of exponents $(\alpha,\beta,z)$ define the growth process, and its universality class \cite{Barabasi95}. Since the universality class is associated with the symmetries, the breaking of the symmetry $h \rightarrow -h$ turns out the KPZ universality class different from that of EW. For example, for the KPZ universality class,  the Galilean invariance~\cite{Kardar86}: 
\begin{equation}
\label{GI}
\alpha+z=2,
\end{equation}
 is a signature of KPZ. 

In this way, the KPZ equation, Eq.~(\ref{KPZ}), is a general nonlinear stochastic differential equation, which can characterize the growth dynamics of many different systems~\cite{Mello01,Reis05,Almeida14,Rodrigues15,Alves16,Carrasco18,Merikoski03,Odor10,Takeuchi13,Almeida17}.
As consequence  most  of these stochastic systems are interconnected.
For instance, the {SS} model~\cite{Krug92,Krug97,Derrida98,Meakin86,Daryaei20}, which is connected with the asymmetric simple exclusion process~\cite{Derrida98}, the six-vertex model~\cite{Meakin86,Gwa92,Vega85}, and the kinetic Ising model~\cite{Meakin86, Plischke87}, all of them are of fundamental importance. It is noteworthy that quantum versions of the KPZ equation have been recently reported that are connected with a Coulomb gas~\cite{Corwin18}, a quantum entanglement growth dynamics with random time and space~\cite{Nahum17}, as well as in infinite temperature spin-spin correlation in the isotropic quantum Heisenberg spin-$1/2$ model~\cite{Ljubotina19, DeNardis19}. 

Despite all effort, finding an analytical,  or even a numerical solution, of the KPZ equation~(\ref{KPZ}) is not an easy task~\cite{Dasgupta96,Dasgupta97,Torres18,Wio10a,Wio17,Rodriguez19} and we are still far from a satisfactory theory for the KPZ equation, which makes it one of the most difficult and exciting problems in modern mathematical physics~\cite{Bertini97,Baik99,Prahofer00,Dotsenko10,Calabrese10,Amir11,Sasamoto10,Doussal16,Hairer13}, and probably one of the most important problem in non equilibrium statistical physics. The outstanding works of Prähofer and Spohn~\cite{Prahofer00} and Johansson~\cite{Johansson00} opened  the possibility of an exact solution for the distributions of the heights fluctuations $f(h, t)$ in the KPZ equation for $1 + 1$ dimensions (for {reviews} see~\cite{Calabrese10,Amir11,Sasamoto10,Doussal16,Hairer13,Johansson00} ).

In a recent work~\cite{GomesFilho21b} the exponents were determined for $2+1$ dimensions using 
\begin{equation}
\label{alf}
\alpha=
\begin{cases}
 1/2 , &\text{ if~~ } d=1\\
\frac{1}{d_f+1}, &\text{ if~~ } d \geqslant 2,\\
\end{cases}
\end{equation}
 where $d_f$ denote the fractal (Hausdorff) dimension of the interface, which has been proved to be as well associated with the global roughness exponent $\alpha$, and the well known result~\cite{Barabasi95}
\begin{equation}
\label{df2}
\alpha=2-d_f,
\end{equation}
for $d=1,2$.  This yields  for $2+1$ dimensions
\begin{equation}
\label{alfn}
z=d_f =  \varphi, \hspace{0.7cm}   \alpha  =  \frac{3-\sqrt{5}}{2}, \hspace{0.7cm}   \beta  =  \sqrt{5}-2,
\end{equation}
where $\varphi=\frac{1+\sqrt{5}}{2}$ is the golden ratio. 

In this work we discuss a fluctuation-dissipation theorem for growth in $d+1$ dimensions and the possible symmetries associated with the fractal geometry of growth.

\section{Fractality, symmetry and universality}

Note that now we do not have just the triad $(\alpha,\beta,z)$ but the quaternary $(d_f,\alpha,\beta,z)$, \textit{i.e.} the fractal dimension and the exponents. 
They are completely connected, thus fractality, symmetry and universality are interconnected as well. Using the  Eqs. (\ref{z}),  (\ref{GI}), (\ref{alf}) and (\ref{df2}) we can determine $(d_f,\alpha,\beta,z)$.
{Therefore} for $2+1$ dimensions, under any point of view, the exponents and fractal dimension have been determined. However, there are questions concerning  the symmetries that we have not even touched.
The first question is why Eq.~(\ref{alf}) has a distinct behavior for $d=1$ and  $d \neq 1$? The value $\alpha=1/2$ for $d=1$, \textit{i.e.} $1+1$ dimensions, is known since the KPZ original work~\cite{Kardar86}. It is a consequence of the validity of the FDT~(\ref{xi}) for this dimension. {Nonetheless,  we have some new elements for $d>1$}, the explicit appearance of this new non-Euclidean dimension requires a more detailed analysis of the involved symmetries.


\section{Fluctuations relations and fractal geometry}

In order to understanding deeply the fluctuation-dissipation relation we have to go back to the works of Einstein, Smoluchowski and Langevin on the Brownian motion~\cite{Brown28,Brown28a,Einstein1905, Einstein56, Langevin08,Vainstein06,Nowak17,Oliveira19}. Langevin proposed a Newton equation of motion for a particle moving in a fluid as~\cite{Langevin08}: 
{
\begin{equation}
\frac{\mathrm{d}P(t)}{\mathrm{d}t}=-\gamma P(t)+f(t),
\label{L}
\end{equation}}
where $P$ is  the particle momentum and $\gamma$ is the friction. The ingenious and elegant proposal was to modulate the complex interactions between  particles, considering all interactions as two main forces:
the first contribution represents a frictional force, $-\gamma P$, where the characteristic time scale is $\tau=\gamma^{-1}$ while the second contribution comes from a stochastic force, $f(t)$, with time scale $\Delta t \ll \tau $, which is related with the random collisions between the particle and the fluid molecules. 
The uncorrelated force $f(t)$ is given by 
\begin{equation} 
\label{FDT}
\langle f(t)f(t')\rangle= 2\gamma \langle P^2 \rangle_{eq} \delta(t-t'),
\end{equation}
where $\langle P^2 \rangle_{eq}=mk_BT$, where $k_B$ is the Boltzmann constant. Note that Eq. (\ref{FDT}) was obtained by imposing that  the mean square momentum
reaches a fixed value given by the equipartition theorem, i.e. there is an energy conservation on the average, which means a time translation symmetry.
Later on Onsager \cite{Onsager31} demonstrated that symmetries in the susceptibility (response functions) were associated with the crystal symmetry.

 More recently, it was observed~\cite{GomesFilho21} that even for $1+1$ dimensions in growth process Eq.~(\ref{FDT}) is not really a FDT, but a relation for the noise intensity, so the FDT relation was completed using the exact result of Krug {\it et al.}~\cite{Krug92,Krug97} for the saturated roughness:
\begin{equation}
\label{Ws}
w_s=\sqrt{\frac{D}{24 \nu} L},
\end{equation}
 in $1+1$ dimensions.  This shows that the saturation is an interplay between noise $D$, { which increases the saturation, and the surface tension $\nu$}, which {opposes} to the curvature, {acting as a ``friction''} for the roughness. {And therefore the FDT for growth can be written as~\cite{GomesFilho21}:} 
\begin{equation}
\label{GFDT}
\left\langle \xi({x},t) \xi({x'},t')\right\rangle = 2b\nu w_s^2\delta({x}-{x'})\delta(t-t'),
\end{equation}
where $b=24/L$.  Moreover, since the noise and the surface tension in the EW equation have their origin in the same flux, the  separation between them  is artificial, consequently,  the  connection is restored. Thus there is no doubt that Eq. (\ref{GFDT}) give us a real fluctuation-dissipation theorem.

For the $d>1$, there is a violation of the FDT for KPZ~\cite{Kardar86,Rodriguez19}, where the Renormalization Group  (RG) approach works for $1+1$ dimensions, but fails for $d+1$,  when $d>1$. 
The violation of the FDT is well-known in the literature, in structural glass~\cite{Grigera99,Ricci-Tersenghi00,Crisanti03,Barrat98,Bellon02,Bellon06}, in proteins~\cite{Hayashi07},   in mesoscopic radioactive heat transfer ~\cite{Perez-Madrid09,Averin10} and as well in ballistic diffusion~\cite{Costa03,Costa06,Lapas07,Lapas08}. Consequently, the place to look for a solution for the KPZ exponents is the FDT for $d+1$ dimensions.

 For a solid, the crystalline symmetries are broken  during the growth process, which creates a interface with a fractal dimension $d_f$~\cite{GomesFilho21}. {Although a numerical solution of KPZ equation was obtained with good precision~\cite{Torres18} for $d=1,2$ and $3$, the exponents can be obtained in an easier way from cellular automata simulations.} For example, the stochastic cellular automaton, etching model~\cite{Mello01,Rodrigues15,Alves16}, which mimics the erosion process by {an} acid has been recently proven to belong to the KPZ universality class~\cite{Gomes19}. Thus, it was used together with the SS model to obtain the fractal dimension and the exponents with a considerable precision~\cite{GomesFilho21b}. 

 Now, we want to discuss how the fluctuation-dissipation theorem is affect by   the interface growth.  In order to do that
let us use the SS model, which is defined as follows. Let $\Omega$ be our $d$-dimensional lattice, at any time $t$:
\begin{enumerate}
    \item Randomly choose a site $i\in{\Omega}$;
    \item If $h(i,t)$ is a local minimum, then $h(i,t+\Delta t)=h(i,t)+2$, with probability $p$;
    \item If $h(i,t)$ is a local maximum, then $h(i,t+\Delta t)=h(i,t)-2$, with probability $q=1-p$.
\end{enumerate}
The above rules generate the SS model dynamics. Let $\eta_{ij}(t)$ denote the height difference between two nearest neighbors sites, \i.e., $ \eta_{ij}(t)=h_i(t)-h_j(t)$. By construction, $\eta_{ij}(t)= \pm 1$. That makes the SS model analytically more treatable and easily associated  to the Ising model. Furthermore, changing the value of $p$ corresponds to changing the value of the tilt mechanism parameter $\lambda$ in KPZ equation. In particular, for $p=q$  the average height is constant, {which} characterizes the EW model. 

\begin{figure}[tbp]
	\begin{center}
 	 \includegraphics[width=1\columnwidth]{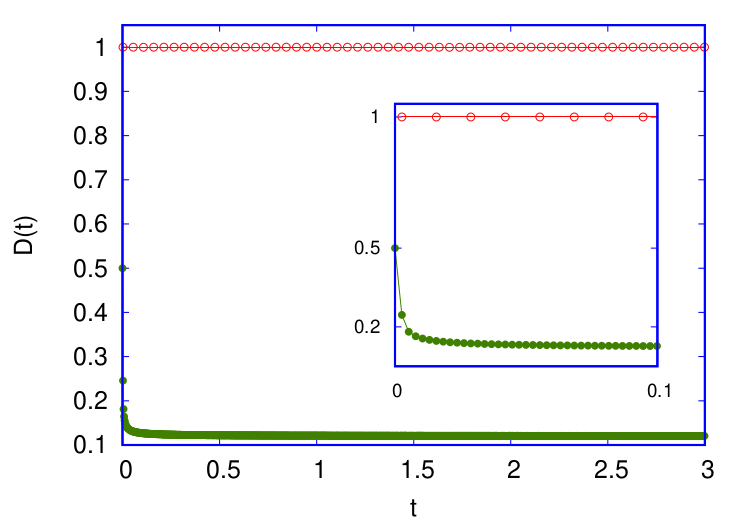}	
	\caption{\label{fig1} Noise intensity as a function of time for the $2+1$ SS model. The upper curve  {corresponds} to the applied noise {while the}  lower curve to the effective noise, i.e.  the noise that actually propagates through the lattice after being filtered by the rules of the SS model. The inset shows the short time behavior.
	}	
	\end{center}
\end{figure}

In Figure~(\ref{fig1}), we show the evolution of the noise intensity  with time, for a $2+1$ SS model. Time is in units of normalized time $t/t_\times$. We use the above rules, periodic boundary conditions, $p=1$, a  $1024 \times 1024$ lattice and we average over $10,000$ experiments.
In the upper curve, we exhibit the  applied  white noise  with mean squared value {equal to} $1$. In the lower curve, we {show} the effective noise, i.e. the remain noise after it has passed through the filter of the rules (2) and (3) above.  {The effective noise intensity depends on the fractality of the interface, decreasing as it is shaped by the SS dynamics, until it stabilizes when the saturation $w(t) \rightarrow w_s$ stabilizes as well.} 

\begin{figure}[tbp]
	\begin{center}
 	 \includegraphics[width=1\columnwidth]{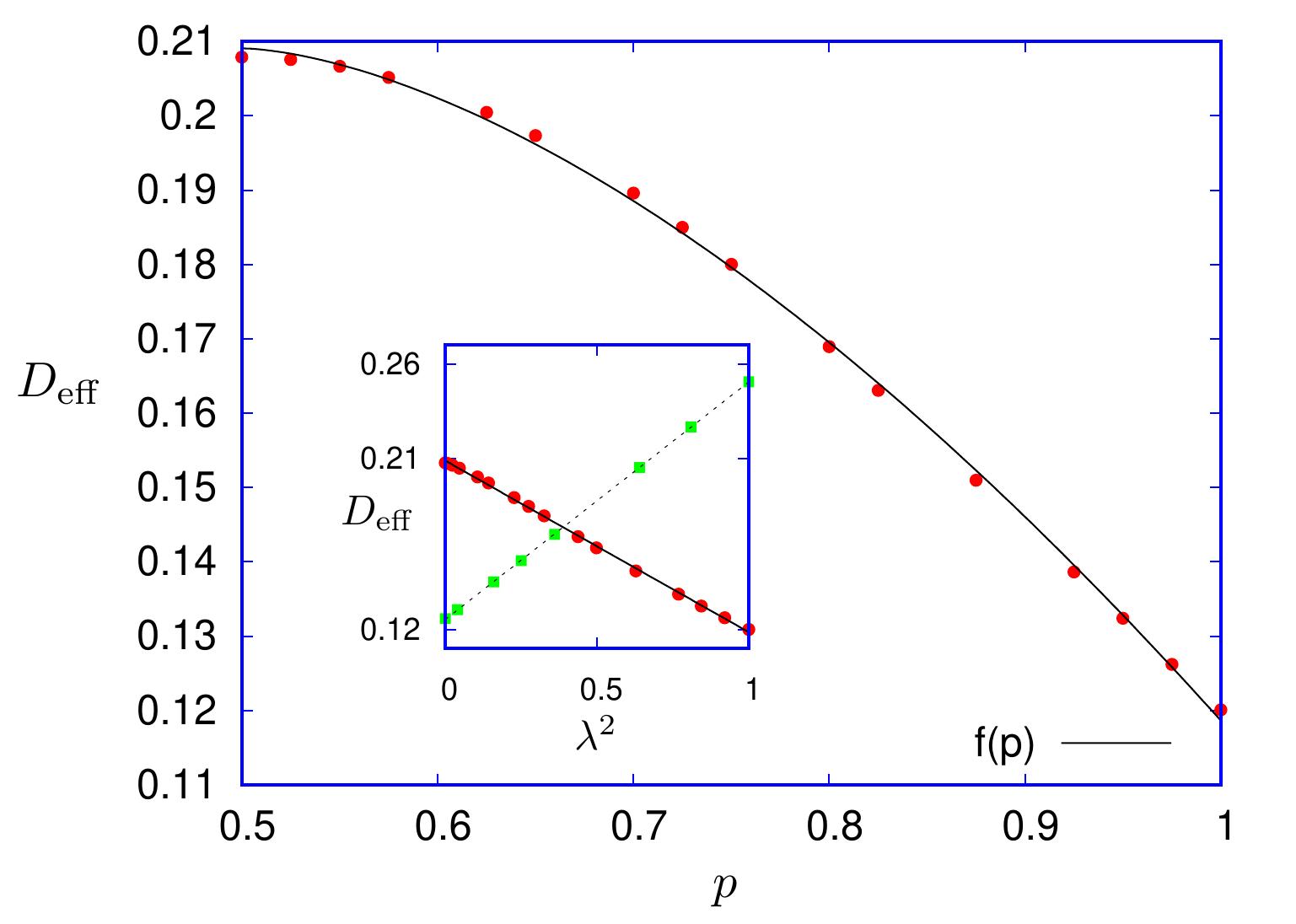}	
	\caption{\label{fig2} Intensity of the effective noise $D_{eff}$  as a function  of  $p$ for  the SS model in $2+1$ dimensions. The continuous curve is the function $f(p)=c_1-c_2(p-q)^\varphi$. Here $\varphi$ is the golden ratio,  $c_1$ and $c_2$ are adjustable constants. In the inset we have $D_{eff}$ as function of $\lambda^2$. The dashed line with positive slope is for $1+1$ dimensions with $\lambda=p-q$ while the other one with negative slope  is for $2+1$ dimensions.}	
	\end{center}
\end{figure}

Let $D_{eff}(t)$ denote the effective noise intensity. In Figure~(\ref{fig2}), we show 
the behavior of $D_{eff}$ as a function of the probability $p$. 
The data points  were obtained from a time average of the noise intensity $D_{eff}(t)$ for each value of $p$  after stabilization (see the lower curve in  Fig.~(\ref{fig1})).
The continuous  curve is the function $f(p)=c_1-c_2(p-q)^\varphi$,  which {adjust} the data very well.
In the inset, we exhibit $D_{eff}$ as function of $\lambda^2$, where, for simplicity, we take the normalized $\lambda \equiv \lambda/\lambda_{max}$. The dashed line with positive slope is for $1+1$ dimensions with 
 $D_{eff} (\lambda)=D^{(1)}_{EW}(1+\lambda^2)$ and $\lambda=p-q$ while the other one is for $2+1$ dimensions with 
 $D_{eff}(\lambda)=D^{(2)}_{EW}+c_3\lambda^2$  and $\lambda = (p-q)^{\varphi/2}$. Here $c_i$ with $i=1,2,3$ and $D^{(d)}_{EW}$, $d=1,2$ are  adjustable constants. The  EW universality class corresponds to the values of both $p$ and $q$ equal to $1/2$ while for $p \neq q$ we have the KPZ universality class.

Up to now, we have not been able to get an analytical proof for the function $f(p)$. However, it shows a direct connection with the fractal geometry of the interface, via the fractal dimension $d_f= \varphi$.

Finally, we can generalize the Eq.~(\ref{Ws})  to obtain a most
general form of $w_s$ as~\cite{GomesFilho21b}:
\begin{equation}
\label{wsg}
w_s= \left( \frac{ D}{b\nu \Phi} L \right)^\alpha,
\end{equation}
and then,  we rewrite the FDT as
\begin{equation}
\label{FDTF}
\left\langle \xi(\vec{x},t) \xi(\vec{x'},t')\right\rangle = 2c\nu w_s^{1/\alpha}\delta^{(d_f)}(\vec{x}-\vec{x'})\delta(t-t').
\end{equation}
Definitions of fractional delta functions can be found in \cite{Jumarie09,Muslih10}. For $1+1$ dimensions $\alpha=1/2$ and Eq. (\ref{FDTF}) reduces to Eq. (\ref{GFDT}). For $d+1$ dimensions with $d \geqslant 2$,   $\alpha$ and $d_f$ is  given by Eq. (\ref{alfn}). Here $c$ is given by
$c=\frac{24 \Phi}{L}$,
where the parameter $\Phi$ is a dimensionless number given by
\begin{equation}
\label{Phi}
\Phi=\left(\frac{D}{\nu} \right)^{\theta/d_f}\left(\frac{\lambda}{\nu} \right)^{\theta},
\end{equation}
with $\theta=0$ for $d=1$, thus $\Phi=1$, and a number close to zero for higher dimensions. For $2+1$ dimensions the etching model yields~\cite{GomesFilho21b} $\Phi=1.00(2)$, for other models $\Phi$ is of the order of unity. It is not necessary that $\Phi$ is of the order of unit, however it sounds strange when a dimensional analysis  has hidden numbers that are too big or too small.  The  Eq.~(\ref{FDTF}) generalizes the result for $1+1$ dimensions~\cite{GomesFilho21}.  This is a step forward, however it should be noted not only that here we have a FDT for each growth equation, but also that the exponent $\alpha$ changes with dimension while the FDT for the Langevin's equation (\ref{FDT}) is very general and independent of the dimension.

Note that Eq. (\ref{GFDT}) reflects a general characteristic of the FDT of being a linear response to small deviation from equilibrium. 
This fact is very clear at saturation, thus the nonlinear term is not so important.  
Note again that going from Eq. (\ref{xi}) to (\ref{GFDT}) does not alter the KPZ equation. Equation (\ref{xi}) is what is applied while Eq. (\ref{GFDT}) is what propagates.\\
\\

\section{The hidden symmetry }
The breaking of the crystalline structure such as discussed above bring us in a first step to a no man's land.
Next we realize that we have universal values,  for example in $2+1$ dimensions $z =d_f=\varphi=1.61803...$, we compare this value  with experiments in electro-chemically induced co-deposition of nanostructured NiW alloy
films, $z=1.6(2)$~\cite{Orrillo17}, dynamics in chemical vapor deposition in silica films, 
 $z=1.6(1)$~\cite{Ojeda00},   on semiconductor polymer deposition $z=1.61(5)$~\cite{Almeida14} and in 
excess of mutational jackpot events in expanding populations
revealed by spatial Luria–Delbrück experiments, $z=1.61$ \cite{Fusco16}. Thus we have agreement with different kinds of recent experiments, and very precise simulations~\cite{GomesFilho21}, it is unlikely to be just a numerical coincidence.
Consequently, there is a well defined fractal geometry for KPZ and the cellular automata associated with it, whose symmetries are unknown.

 The golden ratio $\varphi$ is associated 
with the limit of the Fibonacci  sequence $0,1,1,2,3,5...$ i.e.
a sequence where the element of order n is given
by $F_n=F_{n-1}+F_{n-2}$, thus in the limit $n \rightarrow \infty $ we have the golden ratio
$\varphi= \lim_{n \rightarrow \infty}=F_{n+1}/F_n$. This sequence is very common in growth forms~\cite{Dunlap97}.\\

\textit{Platonic solids and symmetries.\textemdash} The first place to look for symmetry in $3$ dimensions are the platonic solids. 
For example, one can consider the icosahedron or its dual, the dodecahedron, both of these solids have a well-known relationship with the golden ratio.
The icosahedron consists of $20$ identical equilateral triangular faces, $30$ edges and $12$ vertices. 
It has a large group of symmetry and it is isomorphic to the non-abelian group  $A_5$ of all icosahedron rotations \cite{Szajewska14,Hamermesh12}. 
The $A_5$ group has one trivial singlet, two triplets, one quartet, and one quintet. In their matricial representation the generators of the triplets and of the quintet  has elements where $\varphi^k$, with $k=0,1,2$ appears. \\ 
\\ 
\textit{Deterministic fractal cellular automata.\textemdash}
Magnetic systems are probability the best physical system to look for symmetry. Even the most simple Ising hamiltonian system exhibit the symmetric paramagnetic phase, and the symmetric breaking phases (ferro and antiferromagnetic). In addition to the trivial phases,  is possible to  construct additional
fractal symmetry protected topological (FSPT) phases via a decorated defect approach (see \cite{Devakul19} and references there in).

We can define a fractal cellular automaton using the rules of fractal geometry.  For example we can take the set of points  $i=(....-2,-1,0,1,2,...)$ as an infinite line. The rules
\begin{equation}
a^{t+1}_i= a^t_{i-1}+a^t_i+a^t_{i+1}\,\,\, {\text{mod}\,2},
\end{equation}
with the initial condition $a^0_i={\delta_{0i}}$, generate a Fibonacci fractal~\cite{Devakul19}  in the space-time $(i,t)$. From that we get the (Hausdorff) fractal dimension $d_f=1 + \log_2(\varphi) \approx 1.63 $. The golden ratio appears associated with this fractal dimension, however it is not yet the fractal dimension itself.

As a more general example, we can consider the {\it Fibonacci Word Fractal} (FWF). This process is more interesting to us because it is a dynamical growth process and consequently more connect with our physical motion, so we  shall discuss it briefly. FWF are strings over $\{0, 1\}$ defined inductively as follows: $f_0 = 1, f_1 = 0$, $f_n=f_{n-1}f_{n-2}$ (i.e. $f_n$ is the concatenation of the two previous strings, e.g. $f_2=01, f_3=010, \cdots$), for $n\geqslant 2$. The FWF can be associated with a curve using the following drawing rule: for each digit at position $k$, draw a segment forward then if the digit is $1$ stay straight, if the digit is $0$, turn $\theta$ radians to the left, when $k$ is even, or turn $\theta$ radians to the right, when $k$ is odd. For an arbitrary $\theta$, \cite{Monnerot13} shows that
$$d_f= \frac{3\ln \varphi}{\ln(1+\cos \theta + \sqrt{(1+\cos \theta)^2 + 1})}.$$
Note that for an adequate choice of the turning angle $\theta$, we can generate a fractal curve with $d_f=\varphi$. A simple calculation shows that $\theta \approx 0.9902 \frac{\pi}{2}$, i.e almost a  right angle. 

\begin{figure}[tbp]
	\begin{center}
 	 \includegraphics[width=0.9\columnwidth]{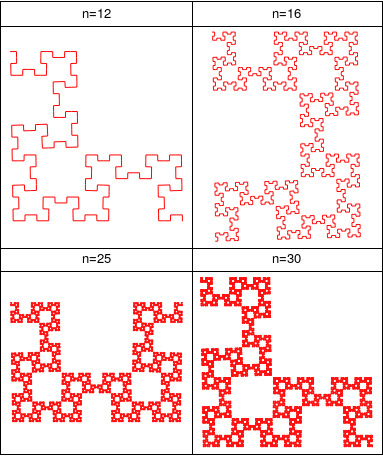}	
	\caption{\label{fwf} The Fibonacci Word Fractal for a turning angle  $\theta \approx 0.9902 \frac{\pi}{2}$. This curve has Hausdorff dimension $\varphi$.  The result is almost indistinguishable from that of $\theta=\pi/2$. However, if we select $n=30$ and we follow the trajectory around the big white square, we see that the cumulative effects are such that the trajectory does not close.
	}	
	\end{center}
\end{figure}

Figure \ref{fwf} shows an example of this fractal for a different number of interactions.

\textit{Stochastic  cellular automata.\textemdash} Now we return to the growth problem described by a stochastic KPZ equation or by a stochastic growth cellular automaton  belonging to the KPZ universality class. There are a number of well-known cellular automata and probably much more to be discovered.   All of them acts in an integer space of dimensions $D_i=d+1$ and generates a fractal space of dimensions $D_f=d_f+1$,  with the same  $d_f$ given by Eq. (\ref{alfn}). The fractal dimension associated with universality via Eq. (\ref{alf}) is itself universal. Therefore, there must be a hidden symmetry and with that new finite non Abelian groups. Is quite natural that we have identified first the groups of perfect crystal,  since we have a visual identification of its properties. For fractal which may present  only average self-similarity it is more hard to find. However, we expect that soon we would be able to unveiling it.  Although we have not yet identified such groups, the discussion above,   particularly the concept of self-similarity, is a starting point for such endeavour. Once it is discovered, its importance will be far beyond KPZ.

\textit{Conclusion.\textemdash} In this work,  we  discuss the  fluctuation-dissipation theorem  for the Kardar-Parisi-Zhang equation in a space of $d+1$ dimensions.  We show how an applied noise  is transformed as it goes through the filter imposed by the rules of a cellular automaton.  In particular we use the SS model, where controlling the probability $p$ we can change the effective noise intensity.  The results support recent work \cite{GomesFilho21b}, which suggest that the effective noise has fractal dimension $d_f$. This fractal dimension is associated with the KPZ exponents from Eq. (\ref{alfn}), in such way that we have now not only the 
triad $(\alpha,\beta,z)$ but the quaternary $(d_f,\alpha,\beta,z)$. This new universality implies that we have a new hidden symmetry for the KPZ universality class. We found a deterministic cellular automaton from which we can control the Hausdorff dimension $d_f$ in such way that, we can obtain $d_f=\varphi$. This may be a starting point for new symmetries relations.

%

\section*{Funding}
This work was supported by the Conselho Nacional de Desenvolvimento Cient\'{i}fico e Tecnol\'{o}gico (CNPq), Grant No.  CNPq-312497/2018-0  and the Funda\c{c}\~ao de Apoio a Pesquisa do Distrito Federal (FAPDF), Grant No. FAPDF- 00193-00000120/2019-79. (F.A.O.).

\bibliography{references}{}

\end{document}